\documentclass[a4,12pt]{article}

\newcommand{\PLA}[3]{Phys.\ Lett.\ A\ {\bf #1},\ #2 (#3)}

\newcommand{\PRL}[3]{Phys.\ Rev.\ Lett.\ {\bf #1},\ #2 (#3)}

\newcommand{\RMAP}[3]{Rep.\ Math.\ Phys.\ {\bf #1},\ #2 (#3)}

\newcommand{\NAT}[3]{Nature\ {\bf #1},\ #2 (#3)}

\newcommand{\PRA}[3]{Phys.\ Rev.\ A\ {\bf #1},\ #2 (#3)}

\newcommand{\JMR}[3]{Jour.\ Mag.\ Res.\ {\bf #1},\ #2 (#3)}
\newcommand{\JPC}[3]{Jour.\ Phys.\ Cond.\ Mat.\ {\bf #1},\ #2 (#3)}

\newcommand{\JPA}[3]{J.\ Phys.\ A:\ Math.\ Gen.\ {\bf #1},\ #2 (#3)}

\newcommand{\EPJD}[3]{Eur.\ Phys.\ J.\ A\ {\bf #1},\ #2 (#3)}

\newcommand{\ga}{\gamma}
\usepackage{graphicx}
\usepackage{setspace}
\usepackage{color}
\title{$2N$ qubit ``mirror states" for optimal quantum communication}
\author{S Muralidharan$^1$ S. Karumanchi$^2$ Sakshi Jain$^3$ 
R. Srikanth$^{4,5}$ and P K Panigrahi$^{6,7}$ \\\\
$^1$ Department of Information and communication technology, \\Royal Institute of Technology(KTH), SE-164 40 Kista, Sweden\\
$^2$ Birla Institute of Technology and Science, Pilani, Rajasthan- 333031, India\\
$^3$ Indian Institute of Technology Bombay, Mumbai, India \\
$^4$ Poornaprajna Institute of Scientific Research, Devanahalli, Bangalore 562 110, India\\
$^5$ Raman Research Institute, Sadashiva Nagar, Bangalore 560012, India\\
$^6$ Indian Institute of Science Education and Research (IISER) Kolkata, Salt Lake,\\ Kolkata - 700106, India\\
$^7$ Physical Research Laboratory, Navrangpura, Ahmedabad - 380 009, India}

\input{Qcircuit}
\begin{document}
\maketitle
\begin{abstract}
We introduce a new genuinely $2N$ qubit state, known as the "mirror state" with interesting entanglement properties.
The well known Bell and the cluster states form a special case of these "mirror states", for $N=1$ and $N=2$ respectively.
It can be experimentally realized using $SWAP$ and multiply controlled phase shift
operations. After establishing the general
conditions for a state to be useful for various communicational protocols involving quantum and classical information, it is shown that the present state can optimally implement algorithms for the quantum teleportation of an arbitrary $N$ qubit state and achieve quantum information splitting in all possible ways. With regard to superdense coding, one can send $2N$ classical bits
by sending only $N$ qubits and consuming $N$ ebits of entanglement. Explicit comparison of the mirror state with the 
rearranged $N$ Bell pairs and the linear cluster states is considered for these quantum protocols. We also show that mirror states are more robust than
the rearranged Bell pairs with respect to a certain class of collisional
decoherence. \\
\end{abstract}
\newpage
\section{Introduction}
Quantum communication protocols such as teleportation \cite{Bennett}, secret sharing \cite{Hillery} and superdense coding \cite{Wiesner} require entangled states. Apart from the regular measures like concurrence \cite{mono} and different types of entropies \cite{mems}, one often needs to characterize entangled states keeping in mind, the nature of the quantum task at hand. It has been observed that the efficacy of a given state for a number of quantum tasks depend not only on the degree of entanglement, but also on "connectedness" and "persistency" \cite{connec}. 
Connectedness refers to the possibility of projecting two qubits of a state into the Bell basis by performing an appropriate local
measurement on the other qubits, while the persistency of entanglement refers to the minimum number of local measurements needed
to completely disentangle the given state. \\
In case of three qubits, the $GHZ$ states are maximally connected, whereas $W$ states \cite{wstate} are not, although the latter has a higher persistency. $GHZ$ states can be used for teleportation and secret sharing, whereas the symmetric $W$ state fails to carry out this task. 
Based on LOCC, often used for quantum communicational tasks, entangled states have been classified only upto four qubits \cite{4q}. Higher dimensional entangled states, not belonging to the existing classifications, have been found through intense numerical search procedures \cite{Brown}, which becomes restrictive as the number of particles increases. Instead, approaches based on symmetry and making use of entangling operations, based on physical Hamiltonians, are often preferred since the same are experimentally feasible. Apart from the generalization of the well known GHZ and W states, an interesting new class of $N$ qubit graph states, known as the cluster states \cite{connec} has been introduced into quantum information theory:
\begin{equation}
|C_N\rangle = \frac{1}{2^{N/2}} \otimes_{a=1}^{N} (|0\rangle_a \sigma_z ^{a+1} + |1\rangle_a), 
\end{equation}
with $\sigma_z^{N+1} = 1$. This state owes its origin to Ising type interactions and it simultaneously exhibits maximal connectedness, with a persistency of entanglement of $\frac{N}{2}$.\\
Quantum teleportation of single and multiqubit states is a field of intense research. In a path breaking work, Bennett \it{et al.} \normalfont \cite{Bennett}, introduced the first scheme for the teleportation of an unknown single qubit state $|\psi_{a}\rangle=\alpha|0\rangle+|\beta|1\rangle$ ($\alpha, \beta \in C,  |\alpha|^2 + |\beta|^2 = 1$), using a two qubit Einstein-Podolsky-Rosen (EPR) pair given by, $|\psi_{\pm}\rangle = \frac{1}{\sqrt{2}} (|00\rangle \pm |11\rangle)_{AB}$ as an entangled resource.  
The same can be achieved using the three qubit GHZ \cite{Karl}, the asymmetric W state \cite{Gorbachev, Agrawal} and the cluster state \cite{Sre1}.
Recently \cite{Lee, Rigo, Yeo, Sreraman}, several schemes have been devised using different types of entangled channels for the teleportation of an arbitrary two qubit state given by $|\psi_2\rangle = \sum_{i_1,i_2}^{1} \alpha_{i_{1}i_{2}}|i_{1}i_{2}\rangle$, where $\alpha_{i_{1}i_{2}} \in C$
and $\Sigma |\alpha_{i_{1}i_{2}}|^2 = 1$. While, there are experimentally feasible states that can teleport single and two qubit states, constructing genuine multiqubit entangled channels which can teleport an arbitrary $N\ (N>2)$ qubit state is a non-trivial task and is of obvious interest to experimentalists. Recently, entangled $2N$ \cite{Chen} and $(2N+1)$ \cite{Man} qubit states have been introduced for this purpose. However, experimental feasibility of these states remains an open question.\\

Entanglement is also crucial to "Quantum secret sharing" which refers to the splitting of
secret information among a group of parties such that none of them can completely
reconstruct it themselves by operating on their own share. Quantum secret sharing is used
for splitting and sharing of both classical and quantum information \cite{Gotit}. Splitting of an arbitrary multiqubit state 
among a group of parties such that none of them can retrieve
the quantum state completely by operating on their own qubits is referred to
as "Quantum Information Splitting (QIS)".
In a landmark paper, Hillery \it{et al.} \normalfont  \cite{Hillery} demonstrated the
 the first scheme for QIS of an unknown single qubit state $|\psi_1\rangle$ among two parties using a three qubit $GHZ$ state as a shared entangled resource. This was also achieved using an asymmetric W state and experimentally realized in
 ion trap systems \cite{Zheng}. QIS of an arbitrary two qubit state
has been carried out using  five \cite{Sre1} and six qubit linear cluster states \cite{jk} respectively. However, both the
GHZ and the cluster states cannot be used for carrying out QIS of arbitrary entangled systems containing more
than two qubits. Here we introduce a new experimentally realizable genuinely entangled $2N$ qubit "mirror state" $|\zeta_{2N}\rangle$ that is different from $GHZ$, $W$ and cluster states under LOCC for $N>2$ and exhibits different entanglement properties for these purposes.  It is given by,
\begin{eqnarray}
|\zeta_{2N}\rangle =\frac{1}{\sqrt{N}} ((\sum_{i_1,i_2...i_N = 0}^{1} R|i_{1}i_{2}...i_{N}\rangle 
\otimes |i_{1}i_{2}...i_{N}\rangle) - 2|1\rangle^{\otimes 2N}).
\end{eqnarray}
Here $R$ is the unitary "Reflection operator", which yields the "mirror image" of the state,
 through the following transformation, $|i_1 i_2 . . .i_N \rangle {\stackrel{R}{\rightarrow}}  |i_N i_{N-1} . . .i_1\rangle$. Owing to this built in reflection symmetry in the state, we call it a "mirror state".

 The paper is organized as follows : in the first section, we define a mirror state and show a way of physically 
 realizing them using $SWAP$ and multiply controlled phase shift operations. After studying the entanglement properties of this
 state, we devise explicit protocols to show that this state can be used for teleporting an arbitrary $N$ qubit state. Then, we establish that the channel capacity of this state reaches the "Holevo bound" and show that $2N$ classical bits can be transmitted with $N$ qubits by using this state as a shared entangled resource. We prove that this scheme is unconditionally secure in
 ideal conditions. In the last section, we compare our state with $N$ Bell pairs, linear cluster states and show that 
 the mirror state is better for the considered quantum communication protocols than the former. 
  
 \section{Physical realization and properties}
 
 To design entangled channels keeping the communication protocols in mind, we found it necessary to implement a series SWAP operations in $N$ Bell pairs as 
\begin{eqnarray}
|\psi_+\rangle_{12}|\psi_+\rangle_{34} . . . |\psi_+\rangle_{(2N-1)2N} \stackrel{SWAP(2,2N){SWAP(4,2N-2)} 
..SWAP(N,N+2)}{\rightarrow} |\zeta_{2N}\rangle^{'}.
\end{eqnarray}
if N is odd and 
\begin{eqnarray}
|\psi_+\rangle_{12}|\psi_+\rangle_{34} . . . |\psi_+\rangle_{(2N-1)2N}\  \stackrel{SWAP(2,2N){SWAP(4,2N-2)} 
..SWAP(N-1,N+3)}{\rightarrow} |\zeta_{2N}\rangle^{'}.
\end{eqnarray}
if N is even, where $SWAP (i,j)$ refers to the Swapping of the $i^{th}$ and the $j^{th}$ qubits respectively. In general, one needs to perform $\left[N/2\right]$ $SWAP$ operations to realize the state. 
The $SWAP$ operation can be realized by switching on the Dzyaloshinskii-Moriya (DM) interaction  \cite{zh} 
in the Heisenberg model between the $i^{th}$ and $j^{th}$ qubits in $N$ Bell pairs, for time $t = \frac{k\pi}{2J}$. $|\zeta_{2N}\rangle$, can be obtained from  $|\zeta_{2N}\rangle^{'}$, by performing a controlled phase shift operation \cite{anil} between the first $N$ qubits. 
These interactions together create $N$ ebits of entanglement between the first $N$ 
and the last $N$ qubits. 
For $N=1$, these interactions takes  $|\psi_{-}\rangle$ to $|\psi_{-}\rangle$ and for $N=2$, 
the $SWAP$ operation between the second and
fourth qubits on two Bell pairs $|\psi_{+}\rangle_{12} \otimes |\psi_{+}\rangle_{34}$
leads to
\begin{equation}
|\zeta_4\rangle^{'} = \frac{1}{2}(|0000\rangle + |0110\rangle + |1001\rangle + |1111\rangle)_{1234}.
\end{equation}
After a controlled phase shift operation on the first two qubits, the state $|\zeta_4\rangle$ reads 
\begin{eqnarray}
|\zeta_4\rangle = \frac{1}{2}(|0000\rangle + |0110\rangle + |1001\rangle - |1111\rangle)_{1234},
\end{eqnarray}
which belongs to the class of cluster states. 
It is well known that the Bell and the cluster states, are
well suited for teleportation of an arbitrary single
and two qubit information respectively. For $N \geq 3$, $|\zeta_{2N}\rangle$ differs from the class of
cluster states  and exhibits different entanglement properties. \\
The state is genuinely entangled according to many measures of entanglement. 
The von-Neumann entropy between the subsystems 
$E(\rho_{1,2...,k}| \rho_{k,...,2N}) = k$; hence, for teleporting an arbitrary $k$ $(k\leq N)$ qubit state, Alice
can have the first $k$ particles and Bob the last $(2N-k)$ particles in $|\zeta_{2N}\rangle$. 
Following Ref. \cite{Mcs},  we can notice that the rank of the reduced density matrix of the $j^{th}$ qubit and the $(2n-j+1)^{th}$ qubit of $|\zeta_{2N}\rangle$, 
\begin{equation}
\rho_{j,2n-j+1} = \frac{I}{4} + \frac{1}{4} (\sigma_{3}\otimes \sigma_{3}) + (\frac{2^{n-1}-2}{2^{n+1}})((\sigma_{1}\otimes \sigma_{1}) - (\sigma_{2}\otimes \sigma_{2}))
\end{equation}
 is two.  This means that it is always possible to project two qubits into a Bell state by performing an appropriate local operations on the other qubits, making $|\zeta_{2N}\rangle$  "maximally connected" like the GHZ and the cluster states. Further, the entanglement of this state persists even after we perform local measurements on the other qubits, making the state "highly persistent". In general one needs to perform a minimum of $N/2$ local measurements to break the entanglement of $|\zeta_{2N}\rangle$ which makes it behave like the cluster state under particle loss.

\section{Quantum communication}
Entanglement can be used for transmitting both classical and quantum information. In this section, we investigate
the usefulness of $|\zeta_{2N}\rangle$ for both of these tasks. 
\subsection{Communicating quantum information}
The general condition for an entangled channel $|\psi\rangle_{AB}$, where
$A$ and $B$ refer to the subsystems of Alice and Bob respectively, to be used for teleportation of
an arbitrary $k$ $(k\leq N)$ qubit state, is that there has to be at least $k$ ebits of entanglement
between them. $|\zeta_{2N}\rangle$, possesses this special property owing to the fact that after we trace out $(2N-k)$ particles, the resulting density matrix is completely mixed. Hence, it can be used 
for the perfect teleportation of an arbitrary $k$ qubit state. \\ We let Alice possess, particles $1$ to $N$ and Bob,
the last $N$ particles. Alice has an arbitrary $N$ qubit state $|\psi_N\rangle$, 
\begin{equation}
|\psi_N\rangle = \sum_{i_1,i_2...i_N = 0}^{1} \alpha_{i_{1}i_{2}...i_{N}}|i_{1}i_{2}...i_{N}\rangle = \sum_{1}^{2^N}\alpha_m|\psi_m\rangle. 
\end{equation}
where $\alpha_{i_{1}i_{2}...i_{N}} \in C$ and $\Sigma|{\alpha_{i_{1}i_{2}...i_{N}}}|^2 = \Sigma|{\alpha_m}|^2 = 1$, which she needs to teleport to Bob.
Alice can perform a $2N$ partite joint measurement on her particles as :
\begin{eqnarray}
|\psi_C\rangle = \sum_{m=1}^{2^N} \alpha_m|\psi_m\rangle  \otimes |\zeta_{2N}\rangle = 
\frac{1}{2^{N/2}} \sum |\phi_{x_{i}}\rangle  U_x (\Sigma \alpha_m|\psi_m\rangle),
\end{eqnarray}
where $|\phi_{x_{i}}\rangle's$ form the orthogonal outcomes of the measurements which,
can be rewritten by making use of the reflection operator as
$|\phi_{x_{i}}\rangle = \sum_l \sum_k (|\psi_l\rangle R |\psi_k\rangle) \ (k\neq l) \ \text{or} \ |\phi_{x_{i}}\rangle = \sum_m (|\psi_m\rangle R |\psi_m\rangle) \ (k=l=m)$. Alice can convey the outcome of her measurement to Bob via $2N$ cbits of information. 
Bob's state collapses to  
\begin{eqnarray}
\sum_k\sum_l (\alpha_k|\psi_l\rangle + \alpha_l|\psi_k\rangle) ( \text {for } \ k \neq l), \nonumber \\ \sum_m \alpha_m|\psi_m\rangle (\text{for \ }k=l=m).
\end{eqnarray}
Bob can obtain $|\psi_{N}\rangle$, by performing an appropriate unitary operation on his particles. 
We now illustrate the working of this protocol for the teleportation of an arbitrary three qubit state using $|\zeta_6\rangle$. The unknown three qubit state, that is to be teleported is given by:
\begin{eqnarray}
\label{eq1}
|\psi_3\rangle_{abc} = (\alpha_1 |000\rangle + \alpha_2 |001\rangle + \alpha_3 |011\rangle + \alpha_4 |111\rangle + 
\alpha_5 |110\rangle + \alpha_6 |101\rangle + \alpha_7 |100\rangle + \alpha_8 |010\rangle)_{abc} . 
\end{eqnarray}
The circuit diagram that generates $|\zeta_6\rangle$ that is used to teleport $|\psi_3\rangle$, is shown in Fig 1 :
 
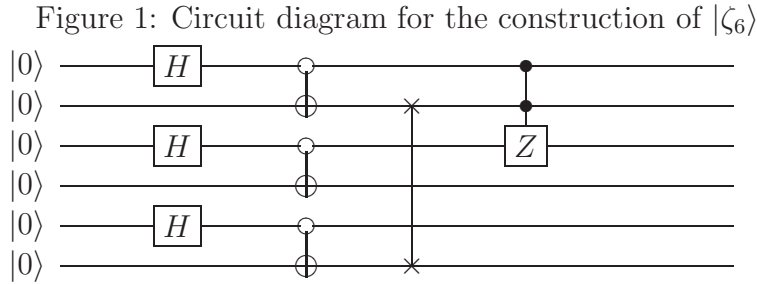
\begin{figure}[h]
	\caption{Circuit diagram for the construction of $|\zeta_6\rangle$}
	\label{fig:CircuitDiagram}
	\leavevmode
\centering	
\Qcircuit @C=3em @R=.3em {
\lstick{\ket{0}}& \gate{H} &\ctrlo{1} & \qw      & \ctrl{2}              & \qw           & \qw \\
\lstick{\ket{0}}& \qw &\targ     & \qswap   & \ctrl{-1}        \qwx      & \qw       & \qw \\
\lstick{\ket{0}}& \gate{H} &\ctrlo{1} & \qw \qwx   & \gate{Z}  \qwx     & \qw     & \qw \\
\lstick{\ket{0}}& \qw &\targ     & \qw \qwx   & \qw  	& \qw      & \qw \\
\lstick{\ket{0}}& \gate{H} &\ctrlo{1} & \qw \qwx   & \qw      & \qw     & \qw \\
\lstick{\ket{0}}& \qw      &\targ     & \qswap \qwx   & \qw        		  & \qw      & \qw \\
}
\end{figure}

The corresponding six qubit "mirror state" could be written as:
\begin{eqnarray}
|\zeta_{6}\rangle = \frac{1}{2\sqrt{2}} (|00\rangle|\psi_+\rangle|00\rangle +|01\rangle|\psi_+\rangle|10\rangle + |11\rangle|\psi_{-}\rangle|11\rangle 
+ |10\rangle|\psi_{+}\rangle|01\rangle)_{163452}
\end{eqnarray}

Alice can perform a joint six partite measurement on "abc163" and classically communicate the outcome of her measurement to Bob via six cbits of information. For instance, if the outcome of Alice's measurement is,  
\begin{eqnarray}
|\phi_x\rangle_2 = \frac{1}{2\sqrt{2}}(|000100\rangle + |001000\rangle + |011111\rangle +  |111110\rangle + 
|100001\rangle + |100011\rangle +\nonumber \\ |101010\rangle + |010101\rangle )_{abc163},
\end{eqnarray}
the corresponding state obtained by Bob is :
\begin{eqnarray}
|\phi_{x_{2}}\rangle = (\alpha_1 |001\rangle + \alpha_2 |000\rangle - \alpha_3 |111\rangle + \alpha_4 |011\rangle
+ \alpha_5 |100\rangle + \alpha_6 |110\rangle + \alpha_7 |010\rangle + \alpha_8 |100\rangle)_{452} .
\end{eqnarray}  
Bob can perform a sutiable controlled phase shift gate and a unitary transformation to
obtain $|\psi_3\rangle$. This completes the teleportation protocol of an arbitrary three qubit state using $|\zeta_{6}\rangle$.
While we note even $N$ Bell pairs can be used for the same task, there are distinct advantages of using the mirror state as
discussed in the last section.
 \subsection{Communicating classical information}
The general condition for an entangled channel $|\Gamma_{AB}\rangle$, where $A$ and $B$ refer to the subsystems of Alice and Bob respectively, to be used for superdense coding of $2k$ cbits $(k\leq N)$ in $k$ qubits is that there has to be at least $k$ ebits of entanglement between $A$ and $B$. The channel capacity of $|\zeta_{2N}\rangle$
reaches the "Holevo bound".  It is interesting to notice that it is always possible to generate a set of
$4^k$ orthogonal states by locally unitary operations on the first $k$ qubits of $|\zeta_{2N}\rangle$.
Hence, if we let, Alice have first $k$ particles and Bob, 
the remaining $(2N-k)$ particles in $|\zeta_{2N}\rangle$, then 
Alice can perform unitary operations on her
particles and convert it into a set of orthogonal states. After performing the unitary operations,
Alice can send her particles to Bob. Bob can then, perform a measurement and retrieve the
classical information. The given entangled channel can be used to send $2k$ cbits by sending $k$ qubits
while consuming $k$ ebits of entanglement. The channel capacity \cite{Bruss} of $|\zeta_{2N}\rangle$ reaches the "Holevo bound", which is given by, $X(\rho^{AB})=N+N-0 = 2N$ allowing $2N$ classical bits to be transmitted
through $N$ quantum bits consuming only $N$ ebits of entanglement. This is impossible using a GHZ, W or the linear 
cluster states of more than five qubits. A detailed comparison follows in the last section.

\section{Quantum secret sharing}
Entanglement can be used for the secret sharing of both quantum and classical information. In the following sections, we will see that the mirror states can be 
used for both these purposes and also its advantages over other entangled states. 
\subsection{Splitting a quantum secret}
Quantum information can be split in more than one way for a given entangled system \cite{Sreramanqis} by distributing 
the qubits between the parties in different ways. The stronger the entanglement of a quantum channel, the more the
number of ways in which quantum information can be split. It has been shown that for a genuinely entangled channel, 
one can split an arbitrary $k$ qubit information in $^{(N-2n)}C_{k-2}$ ways. By substituting $k=2$ in this it can be seen that through a genuinely entangled channel of $N$ qubits, a maximum of $(N-2n)$ protocols can be devised for the 
QIS of an arbitrary $n$ qubit state among two parties in the case where they need not meet \cite{Sreramanqis}. According to this theorem, one can devise a maximum of $k$ protocols for QIS of an arbitrary $(N-k)$ state $(k<N)$ using  $|\zeta_{2N}\rangle $. 
A protocol can be considered successful only if, after Alice performs the measurement, Bob-Charlie
system collapses into a partially entangled state. 
In general any $P$ qubit entangled state having $P\geq(2N+1)$ qubits can be used for QIS of an arbitrary $N$ qubit state, if it is possible to project any $(2N-1)$ qubits into the form
\begin{equation}
\sum_{1}^{2^{( N-1)}} |\Omega_i\rangle_{N-1} U_1\otimes U_2 \otimes...\otimes U_N  |\rho_i\rangle_N
\end{equation}
by performing local measurements on the other qubits. Here $|\rho_i\rangle_N$ refers to the Bell and the GHZ states
for $N=2$ and $N>2$ respectively, $|\Omega_i\rangle$ refers to the computational basis which contains one qubit less than $|\rho_i\rangle_N$ and $U_i \in (\sigma_1, I)$ represents a bit flip or an identity operation which acts
on $|\rho_i\rangle_{N}$ rendering another orthogonal Bell or GHZ state. 
While there are a total of $2^N$ orthogonal GHZ states for a $N$ qubit system, we consider
only half of them, namely the ones with a positive phase difference between the superposition terms. This condition is satisfied for
$\zeta_{2N}$ and it can be used for the QIS of an arbitrary $k$ qubit state for $k<N$.

Let us consider the example of $|\zeta_6\rangle$ for splitting up of $|\psi_{2}\rangle$, in the case where Alice possesses the first three qubits, Bob possesses the fourth qubit and Charlie possesses the last two qubits. Alice can perform a joint five qubit measurement on her qubits and convey
the outcome of her measurement to Charlie. For instance,
if the outcome of Alice's measurement reads $\frac{1}{\sqrt{2}}(|0\rangle|\psi_{+}\rangle00\rangle + |1\rangle|\psi_{+}\rangle11\rangle)$
then the Bob-Charlie system collapses to the entangled state, $(\alpha_{00}|000\rangle - \alpha_{01}|111\rangle + \alpha_{10}|001\rangle + \alpha_{11}|110\rangle)$.
Bob can now perform a measurement, in the basis $\frac{1}{\sqrt{2}} (|0\rangle \pm |1\rangle)$, and 
communicate the outcome of his measurement to Charlie. Having known the outcomes of both their measurements, 
Charlie obtains the state by performing an appropriate controlled phase shift gate followed by an unitary operation. 
Hence, this protocol succeeds. All the protocols succeed for the QIS of $|\psi_1\rangle$ and $|\psi_2\rangle$, using $|\zeta_{6}\rangle$ as an entangled channel. This is however not the case if we use two $GHZ$ or three Bell pairs
as entangled channels. A detailed comparison follows in section 5.  

\subsection{Splitting of classical information}
Hillery \it {et al. }\normalfont \cite{Hillery} presented another scheme by which a random classical bit string could be securely shared among a group of dealers. Considering the fact that the probability distributions of all secret bits should be equal and the scheme should be robust against eavesdropping attacks, It was recently shown that only quantum states which are distillable for all bipartite splits of the number of parties involved can be used for this purpose \cite{kola}. As we will see in the next section, mirror states satisfy this condition, while the rearranged Bell pairs dont. Hence, "mirror states" have an edge over rearranged Bell pairs for sharing of classical information. 

\section{Decoherence and error correcting properties}

Decoherence occurs because quantum systems are in general not 
strictly isolated but open, interacting with an environment, usually
modelled as a bath of qubits or oscillators. 
Its action can be described as ${\cal D}(\rho) = \sum_k E_k \rho E_k^\dag$,
where $E_k$ are Kraus operators, satisfying the completeness
condition $\sum_k E^\dag_k E_k=I$ \cite{nielsen}. Different forms of
interactions give rise to different noise
processes. A non-dissipative interaction of a quantum system with the
environment 
results in a phase damping channel \cite{bs}. A dissipative interaction
with a squeezed thermal bath gives rise to the squeezed generalized
amplitude damping channel \cite{sb}, which generalizes the 
generalized amplitude damping channel \cite{nielsen}. It is well known that because the Pauli operators form a
(unitary) operator basis for all complex $2 \times 2$ matrices,
the un-normalized state $E_j|\psi\rangle$ due to noise acting on
some given qubit $k$, can be written as a
superposition of the four terms $|\psi\rangle, X_k|\psi\rangle, 
Z_k|\psi\rangle$ and $X_kZ_k|\psi\rangle$. A quantum 
error correcting code (QECC) is a Hilbert subspace, such that
each of these errors drives any element in it to a distinct,
orthogonal subspace.

More generally, let $P$ be the projector onto a QECC $C$. Then $C$ can
correct the error set $\{E_j\}$ if and only if the following error correcting
conditions are satisfied:
\begin{equation}
\label{qecc}
PE^\dag_jE_kP = \alpha_{jk}P,
\end{equation}
where $\alpha \equiv \{\alpha_{jk}\}$ is some Hermitian matrix \cite{kl97}.
As noted earlier, $|\zeta_{2N}\rangle$ satisfies the property
that of the $4^k$ single-qubit Pauli operators belonging
to $\Xi \equiv \{I,X,Y,Z\}^{\otimes k}$ $(k \le N)$ take $|\zeta_{2N}\rangle$
to orthogonal states. Therefore, the mirror states satisfy
the error correcting condition for up to $N$ arbitrary single qubit
errors, with $\alpha_{jk} = \delta_{jk}$,
and are seen to possess an intrinsic error correcting property,
a feature that enhances their value in the above communication protocols. 
The error syndrome is obtained by measurement in the `mirror state basis'.

The above properties of the mirror states do not necessarily set them
apart from rearranged Bell pairs with respect to decoherence, but
a comparison of collisional decoherence properties 
does It has been shown that all decoherence maps of qubits, can be modelled as sequences of collisions between the system qubits under consideration and qubits from their environment \cite{Buzek1, Buzek2} . Motivated by this, we compare the effects of collisional decoherence on $|\zeta\rangle_4$ and $|\zeta'\rangle_4$. 
Following \cite{Yeo1}, we consider a system of $N$ qubits, $A_1, \cdots, A_N$, initially decoupled from an environment, i.e., the initial system-environment state is given by $\rho_{A_1\cdots A_N} \otimes \Xi_{env}$. The environment can be modelled as a set of $\prod^N_{i = 1}N'_i$ qubits: $E_{ij}$, with $1 \leq i \leq N$ and $1 \leq j \leq N'_i$.  These ``environment qubits" are initially in a factorized state: $\Xi_{env} = \bigotimes^N_{i = 1}\bigotimes^{N'_i}_{j = 1}\xi_{E_{ij}}$, with $\xi_{E_{ij}} = \xi$ for all $E_{ij}$.  They do not interact between themselves, and each $E_{ij}$ undergoes a bipartite collision with each system qubit $A_i$ only once. This gives rise to local dephasing channels. For one qubit case, the master equation for the dephasing channel is given by \cite{Yeo1},
\begin{equation}
\frac{d}{dt}\rho_{A_i}(t) = -i\frac{\phi}{2}[\sigma^3_{A_i}, \rho_{A_i}(t)] - \frac{1}{2}ln\lambda_i[\sigma^3_{A_i}\rho_{A_i}(t)\sigma^3_{A_i} - \rho_{A_3}(t)].
\end{equation}	
assuming, $\lambda_{ij} = \lambda_i$ and $\phi_{ij} = \phi_{i}$ for all j and $0 < \lambda_{ij} < 1$\\

If a single qubit system is expressed as,
\begin{equation}
\rho_{A_i} = \rho^{00}_{A_i}P^0_{A_i} + \rho^{01}_{A_i}S^+_{A_i} + \rho^{10}_{A_i}S^-_{A_i} + \rho^{11}_{A_i}P^1_{A_i},
\end{equation}
where
$$
P^0 \equiv |0\rangle\langle 0|, 
P^1 \equiv |1\rangle\langle 1|, 
$$
\begin{equation}
S^+ \equiv |0\rangle\langle 1|,
S^- \equiv |1\rangle\langle 0|.
\end{equation}
To obtain $\rho'_{A_1...A_N}$, the state after $\rho_{A_1...A_N}$ passes through the decoherence channels, which can be treated as a series of collisions, we note that after the $k_i$th collision, the state evolves as follows;
\begin{eqnarray}
P^{0, 1}_{A_i} \longrightarrow P^{0, 1}_{A_i}\\ \nonumber
S^{\pm}_{A_i} \longrightarrow \gamma_i\exp(\pm i\Phi_i)S^{\pm}_{A_i},
\end{eqnarray}
where
\begin{eqnarray}
\gamma_i \equiv \prod^{k_i}_{j = 1}\lambda_{ij},\ \Phi_i \equiv \sum^{k_i}_{j = 1}\phi_{ij}.\\ \nonumber
0 \leq \lambda_i \leq 1
\end{eqnarray}

We studied the robustness of $|\zeta\rangle_4$ and $|\zeta'\rangle_4$ by analysing the negativities of $\rho'_{A_1...A_4}$ \cite{Vidal1}. A necessary condition for $N$-party distillability is that the partial transpositions with respect to any group of parties are nonpositive. Negativity equals the absolute value of the sum of the negative eigen values of the transposed matrix \cite{Peres}. N-party distillability implies non-zero negativities. 

Our results for the negativity of the rearranged Bell state are as follows:
\begin{eqnarray}
{\cal N}[\rho'^{B}_{(A_1)A_2A_3A_4}] &=& \frac{1}{2}(\ga_1\ga_4)\\ \nonumber
{\cal N}[\rho'^{B}_{A_1(A_2)A_3A_4}] &=& \frac{1}{2}(\ga_2\ga_3)\\ \nonumber
{\cal N}[\rho'^{B}_{A_1A_2(A_3)A_4}] &=& \frac{1}{2}(\ga_2\ga_3)\\ \nonumber
{\cal N}[\rho'^{B}_{A_1A_2A_3(A_4)}] &=& \frac{1}{2}(\ga_1\ga_4)\\ \nonumber
{\cal N}[\rho'^{B}_{(A_1A_2)A_3A_4}] &=& \frac{1}{2}(\ga_1\ga_2\ga_3\ga_4 + \ga_1\ga_4 + \ga_2\ga_3)\\ \nonumber
{\cal N}[\rho'^{B}_{(A_1)A_2(A_3)A_4}] &=& \frac{1}{2}(\ga_1\ga_2\ga_3\ga_4 + \ga_1\ga_4 + \ga_2\ga_3)\\ \nonumber
{\cal N}[\rho'^{B}_{(A_1)A_2A_3(A_4)}] &=& 0
\end{eqnarray}
Thus, $\ga_{crit} = 0$, below which, the rearranged Bell state does not remain 4 party distillable entangled.

The corresponding calculations for the 4 qubit mirror state yields:
\begin{eqnarray}
{\cal N}[\rho'^{M}_{(A_1)A_2A_3A_4}] &=& \frac{1}{2}(\ga_1\ga_4)\\ \nonumber
{\cal N}[\rho'^{M}_{A_1(A_2)A_3A_4}] &=& \frac{1}{2}(\ga_2\ga_3)\\ \nonumber
{\cal N}[\rho'^{M}_{A_1A_2(A_3)A_4}] &=& \frac{1}{2}(\ga_2\ga_3)\\ \nonumber
{\cal N}[\rho'^{M}_{A_1A_2A_3(A_4)}] &=& \frac{1}{2}(\ga_1\ga_4)\\ \nonumber
{\cal N}[\rho'^{M}_{(A_1A_2)A_3A_4}] &=& \frac{1}{2}(\ga_1\ga_2\ga_3\ga_4 + \ga_1\ga_4 + \ga_2\ga_3)\\ \nonumber
{\cal N}[\rho'^{M}_{(A_1)A_2(A_3)A_4}] &=& \frac{1}{2}(\ga_1\ga_2\ga_3\ga_4 + \ga_1\ga_4 + \ga_2\ga_3)\\ \nonumber
{\cal N}[\rho'^{M}_{(A_1)A_2A_3(A_4)}] &=& \max(\frac{1}{4}(\ga_1\ga_2\ga_3\ga_4 + \ga_1\ga_4 + \ga_2\ga_3 - 1), 0)
\end{eqnarray} 
Thus, assuming all $\ga_i$ to be equal, we obtain $\ga_{\rm crit} = -1 + \sqrt{2}$. Thus, there exists non-negative values of $\ga$, for which  $|\zeta\rangle_4$ is distillable entangled as against $|\zeta'\rangle_4$. The following argument could be easily extended for N qubits, proving that the mirror states are inherently robust compared to rearranged Bell pairs.

\section{Comparison}
In this section, we compare the mirror state with well known entangled states in quantum information theory for quantum communication tasks.
\subsection{N Bell pairs}
In $|\zeta_{2N}\rangle^{'}$ which is essentially $N$ Bell pairs rearranged, there are $N$ ebits between the first and last $N$ qubits respectively. Hence, even $|\zeta_{2N}\rangle^{'}$ can be used for the teleportation of $|\psi_N\rangle$ between two parties. 
However, this teleportation scheme is restricted because Alice and Bob should possess only the first and last $N$ qubits
of $|\zeta_{2N}\rangle'$ respectively. Any other distribution of qubits will fail to achieve the purpose. Whereas, optimal teleportation will work out with many such combinations  using $|\zeta_{2N}\rangle$ as an entangled channel.
In the case of QIS, not all $(N-k)$ protocols work out for the QIS of an arbitrary $k$ qubit state among two parties using
$|\zeta_{2N}\rangle'$, whereas all the protocols work out using $|\zeta_{2N}\rangle$ as a
shared entangled resource. This could be illustrated with the following example : Let us consider
$|\zeta_{2N}\rangle^\prime$ for the QIS of an arbitrary two qubit system. We let Alice, 
Bob and Charlie possess two particles each. Initially, Alice 
performs a joint four partite measurement on the arbitrary two qubit state $|\psi_2\rangle$ and
her part of the entangled state and conveys its outcome to Charlie via four classical bits. 
Now, the Bob-Charlie system, then collapses into a product state thereby the
protocol fails. This happens because, there is not enough entanglement 
between the Bob-Charlie system. However, the same protocol would have worked out
had we used the mirror state $|\zeta_{2N}\rangle$ instead of $|\zeta_{2N}\rangle^{'}$, which 
reflects that the mirror state is more entangled than groups of Bell pairs suitably rearranged and 
can hence carry out tasks that the former cannot. Further, rearranged Bell pairs cannot be used for
splitting classical information as they are prone to certain eavesdropping attacks. While, the mirror states
are useful for this purpose.

Apart from this fact, another interesting observation, following
from the discussion in previous section, is that
mirror states are more robust against collisional decoherence
that rearranged Bell pairs, making mirror states more viable from
the prespective of application.

\subsection{Cluster states}
In the case of a $2N$ qubit linear cluster states $|C_{2N}\rangle$, it can be seen that there are no $N$ entangled bits between
any two subsystems respectively for $N\geq 3 $ \cite{Lu}. Hence, it cannot be used for the teleportation of an arbitrary $N$ qubit
state and for QIS of an arbitrary $k$ qubit state $|\psi_k\rangle$ for $k>2$. However, it is well known that linear cluster states have only limited power as they can be classically simulated \cite{classical}. A detailed comparison of the entanglement properties of the $2D$ cluster states and the mirror states will soon follow. 

\section{Conclusion} 
In conclusion, we introduced a new genuinely entangled $N$ partite
analogue of the Bell state known as the "mirror state" with spectacular properties. The proposed state is experimentally realizable by performing several
$SWAP$ operations between various qubits, followed by a controlled phase shift operation between the first $N$ qubits
in $N$ Bell pairs. Since $SWAP$ and controlled phase shift operations have been experimentally
realized in different systems, the production of the state is experimentally accessible.
This state turns out to be an important resource for quantum communication purposes like teleportation, information splitting and superdense coding. The introduced "mirror state" equals the well known Bell and the cluster states for $N=1$ and $N=2$ and differs from the class
of cluster states for $N\geq3$. It is shown that, the proposed state can be used for the teleportation of
an arbitrary $N$ qubit state and information splitting of an arbitrary $(N-k)$ qubit state. 
The state is found to be an excellent resource for superdense coding as well. The given entangled channel can also be used to send
$2k$ cbits by sending $k$ qubits by utilizing $k$ ebits of entanglement (k $\leq$ N), making the superdense coding capacity reach
the "Holevo bound". In general, one can generate a larger dimensional mirror state from two smaller dimensional mirror states. This along with other alternatives for the generation of mirror states are under current investigation. The experimental creation of these states, is still a challenge although its experimentally feasible in
condensed matter and NMR systems. We have also proved that the mirror states possess better distillability properties compared to the rearranged Bell pairs under the collisional decoherence model, thereby making it a better choice for quantum communication protocols. Apart from the quantum communicational protocols discussed here, we hope that the present state with useful entanglement properties will find applications in other aspects of quantum information science.

\end{document}